\documentclass[cameraready]{Interspeech}
\usepackage{multirow}
\title{MSR-Codec: A Low-Bitrate Multi-Stream Residual Codec for High-Fidelity Speech Generation with Information Disentanglement}

\author[affiliation={1}]{Jingyu}{Li}
\author[affiliation={1}]{Guangyan}{Zhang}
\author[affiliation={2}]{Zhen}{Ye}
\author[affiliation={3}, correspondingauthor]{Yiwen}{Guo}
\address{
    $^1$ LIGHTSPEED, Hong Kong SAR \\
    $^2$ The Hong Kong University of Science and Technology, Hong Kong SAR \\
    $^3$ Independent Researcher
}

\email{\{lijingyu0125,gyzhang\}@link.cuhk.edu.hk, zhenye213@gmail.com, guoyiwen89@gmail.com}


\usepackage{comment}

\begin{document}

\maketitle

\begin{abstract}
Audio codecs are a critical component of modern speech generation systems. This paper introduces a low-bitrate, multi-scale residual codec that encodes speech into four distinct streams: semantic, timbre, prosody, and residual. This architecture achieves high-fidelity speech reconstruction at competitive low bitrates while demonstrating an inherent ability for information disentanglement. We construct a two-stage language model for text-to-speech (TTS) synthesis using this codec, which, despite its lightweight design and minimal data requirements, achieves a state-of-the-art Word Error Rate (WER) and superior speaker similarity compared to several larger models. Furthermore, the codec's design proves highly effective for voice conversion, enabling independent manipulation of speaker timbre and prosody. Our inference code, pre-trained models, and audio samples are available at \url{https://github.com/herbertLJY/MSRCodec}.
\end{abstract}

\section{Introduction}
\label{sec:intro}
With the fast development of large language models (LLMs), LLM-based systems have become the mainstream approaches in the task of text-to-speech (TTS) and achieved state-of-the-art performance. Central to these systems is the \textbf{neural audio codec}~\cite{wang2023neural,chen2024vall,du2024cosyvoice2,ye2025llasa,guo2024fireredtts}, a critical module that transforms continuous speech waveforms into sequences of discrete tokens. In a typical generation pipeline, an LLM predicts these speech tokens conditioned on input text, and a corresponding decoder then synthesizes the final audio from this token sequence.

A primary goal for modern neural codecs is to achieve high compression rates without sacrificing quality. Models like SoundStream~\cite{zeghidour2021soundstream} and Encodec~\cite{DBLP:journals/tmlr/DefossezCSA23} use stacked convolutions to compress speech by over 100 times, using Residual Vector Quantization (RVQ) to maintain high-fidelity reconstruction. However, this fidelity often comes at the cost of a high bitrate—the amount of data required to store the tokenized speech. For instance, popular codecs may require bitrates of 6kbps or more, which increases the complexity for storage, transmission, and the downstream language models that must predict these longer token sequences.

To address this efficiency challenge, researchers have explored several information disentanglement strategies~\cite{ren2024fewer,li2024single}. One prominent approach separates speech into time-variant frame-level tokens and time-invariant codes that capture global information, such as speaker timbre and environmental sounds. As this global information does not scale with the speech's length, this method can significantly reduce the required bitrate. For example, LSCodec~\cite{guo25_interspeech} effectively disentangles timbre by explicitly encoding speaker information with a prompt network. A more structured approach, seen in models like SemantiCodec~\cite{liu2024semanticodec}, factorizes speech into its fundamental components: semantic information (linguistic content) and acoustic information. Semantic tokens, which can be extracted via supervised ASR tasks~\cite{du2024cosyvoice} or self-supervised models like Hubert~\cite{hsu2021hubert}, capture what is being said but are insufficient on their own for reconstruction. They must be complemented by another set of acoustic tokens or a generative model to synthesize the final high-fidelity speech.

In this work, we propose a highly efficient, low-bitrate codec that advances this disentanglement paradigm for speech generation. Inspired by factorized representations, we encode speech into four distinct streams: \textbf{semantic}, \textbf{timbre}, \textbf{prosody}, and a fine-grained \textbf{residual} stream. We leverage pre-trained models to extract semantic (Hubert) and timbre (speaker embeddings) information. Our core novelty lies in how these streams are integrated. While prior work~\cite{ju2024naturalspeech} relies on complex mechanisms like adversarial training to enforce disentanglement, our model achieves this implicitly through a simple and stable cascaded architecture. Information from each stream is progressively fused via residual connections, naturally encouraging each component to capture distinct attributes of the speech signal. By employing multi-scale processing and smaller codebooks, our codec achieves a compression factor of over 200x (62.5 tokens per second) while maintaining high-quality output. We construct a two-stage LLM for TTS synthesis using our codec, demonstrating that this disentangled representation leads to superior performance. Our lightweight system achieves a state-of-the-art Word Error Rate (WER) and superior speaker similarity. Notably, it outperforms several larger, state-of-the-art models while being trained on a fraction of the data and generating speech significantly faster. Furthermore, the codec's inherent disentanglement enables highly effective zero-shot voice conversion of both timbre and prosody, confirming the successful separation of these key speech factors.

\begin{figure*}[t]
  \centering
  \includegraphics[width=0.95\linewidth]{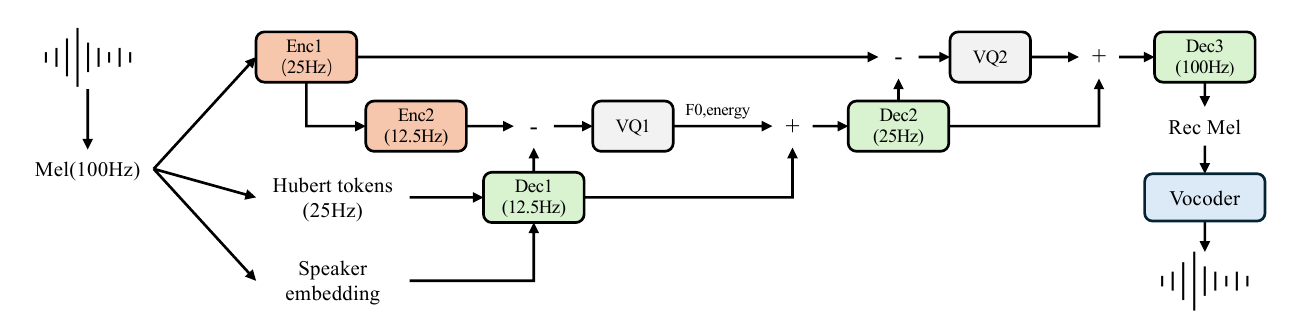}
   % \vspace{-4mm}
  \caption{The overview structure of the codec, the output frame rate of each module is presented in the figure}
  \label{fig:codec}
  % \vspace{-2mm}
\end{figure*}

\section{Method}
\label{sec:method}

Our proposed method consists of two primary components. First, we introduce a \textbf{Multi-Stream Residual Codec} that disentangles speech into four distinct streams and encodes them at a low bitrate. Second, we present a \textbf{Two-Stage TTS Model} designed to generate the tokens for this codec from input text.

\subsection{The Structure of Codec}
\label{ssec:codec}
The codec transforms an input Mel-spectrogram into a highly compressed, factorized representation and then reconstructs it. As illustrated in \autoref{fig:codec}, it processes speech into four streams: timbre, semantics, prosody, and a final residual stream. These are progressively fused in a cascaded architecture, and a pre-trained vocoder converts the final Mel-spectrogram into a waveform.

\subsubsection{Base Layer: Timbre and Semantic Streams}
The foundation of our representation is formed by the speaker's identity and the linguistic contents.

\textbf{Timbre Information:} We use a pre-trained CAM++~\cite{wang23ha_interspeech} network to extract a single, time-invariant speaker embedding for each utterance, which is $l2$-normalized before feeding into the codec.

\textbf{Semantic Information:} To represent the linguistic content, we use a pre-trained Hubert model to extract features at a 25Hz frame rate, which are then quantized into discrete semantic tokens.

The semantic tokens and speaker embedding are fused together using Dec1. The resulting features are downsampled to a 12.5Hz sequence, forming a base representation of the speech's content and speaker identity. Both the speaker and Hubert encoders remain frozen during training.

\subsubsection{Prosody Streams}

While the base layer captures what is said and by whom, it lacks prosodic detail such as pitch and rhythm. We model this at a 12.5Hz frame rate, which is well-suited for these slower-moving contours.

\textbf{Residual Prosody Capture:} To generate the features for this stream, the input Mel-spectrogram is passed through a sequence of two encoders. First, Enc1 transforms the Mel-spectrogram into a 25Hz feature map. This output is then passed through Enc2, which further downsamples it to a 12.5Hz feature map. The prosody information is captured in the residual domain as the difference between this Enc2 output and the output of Dec1. This difference is then quantized by a single codebook (VQ1) to generate prosody tokens.

\textbf{Explicit Supervision}: To ensure these tokens specifically learn prosodic attributes, the quantized features are explicitly trained to predict the corresponding frame-level $F_0$ and spectral energy via an auxiliary predictor.

\textbf{Fusion}: The resulting prosody features are then element-wise added to the base timbre-semantic stream and passed to the next decoder stage (Dec2) for further fusion.

\subsubsection{Residual Stream}
Finally, to ensure high-fidelity reconstruction, we capture fine-grained acoustic details, such as environmental factors and complex textures, in a residual stream. A higher temporal resolution is necessary for this task.

\textbf{Acoustic Detail Capture:} This stream utilizes the 25Hz feature map produced by Enc1. The residual information is calculated as the difference between Enc1's output and the upsampled output from the prosody-enhanced stream (from Dec2). This difference is then quantized using a second codebook (VQ2) to create residual tokens.

\textbf{Final Reconstruction:} These 25Hz residual features are element-wise added to the preceding stream and passed to the final decoder (Dec3) to reconstruct the Mel-spectrogram.

\subsection{Training Objectives}
\label{ssec:loss}
The codec is trained with a combination of three loss functions:

\textbf{Reconstruction Loss ($L_{recon}$)}: This is the primary objective, calculated as the sum of $\ell_1$ and $\ell_2$ distances between the ground-truth and reconstructed Mel-spectrograms,
\begin{equation}
  L_{recon} = \frac{1}{N}\sum_{n=i}^{N}(||x,\hat{x}||_1+||x,\hat{x}||_2)
  \label{eq:L_recon}
\end{equation}
where $x$ is the input and $\hat{x}$ is the reconstructed one. We also apply this loss to the intermediate outputs of Dec1 and Dec2 to stabilize training and ensure each stage contributes effectively to the reconstruction.

\textbf{Adversarial Loss ($L_{adv}$):} A discriminator is trained to distinguish between real ($x$) and reconstructed spectrograms ($\hat{x}$), improving the perceptual quality and naturalness of the output.

\textbf{Prosody Loss ($L_{prosody}$):} An MSE loss is applied between the predicted prosody outputs from VQ1 and the target $F_0$ and energy values, encouraging the prosody stream to capture this specific information.

\subsection{The Structure of the TTS Model}
\label{ssec:tts}

\begin{figure}[t]
  \centering
  \includegraphics[width=\linewidth]{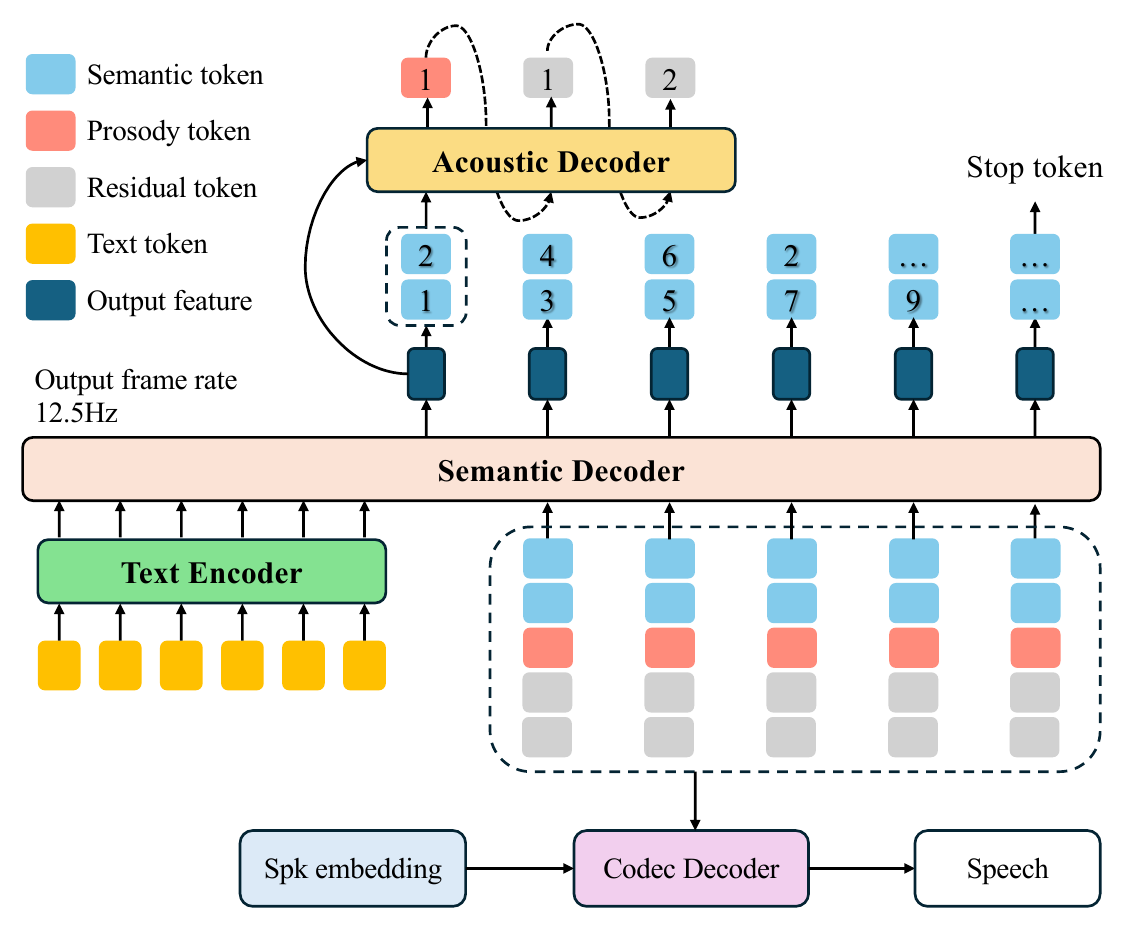}
  % \vspace{-2mm}
  \caption{The network structure of the TTS model. The index denotes the order of each token. Best viewed in colors}
  \label{fig:tts}
  % \vspace{-3mm}
\end{figure}

We build a TTS model to generate speech using the proposed codec, as shown in  ~\autoref{fig:tts}. The architecture consists of a text encoder and a two-stage autoregressive language model.

\textbf{Stage 1: Semantic Prediction with the Semantic Decoder:} The primary component is an autoregressive Semantic Decoder, which operates on features from the text encoder. At each step, the Semantic Decoder generates an output feature at a 12.5Hz rate. A split classification head then predicts two 25Hz semantic tokens from this single feature.

\textbf{Stage 2: Acoustic Prediction with the Acoustic Decoder:} A smaller, secondary Acoustic Decoder is responsible for predicting the acoustic details for each frame. It takes the output feature from the Semantic Decoder and the newly predicted semantic tokens as input to generate the corresponding prosody and residual tokens for that time step.

\textbf{Autoregressive Generation:} The generation process proceeds autoregressively at 12.5Hz. The input to the next step of the primary Semantic Decoder is formed by concatenating its own output feature from the current step with the embeddings of all tokens generated in that step (semantic, prosody, and residual). Generation terminates when a special stop token is predicted by the Semantic Decoder. The speaker embedding is not used during token prediction; it is only provided as a condition to the codec's decoder during the final synthesis stage.

% We build a TTS model for speech generation with the proposed codec. The structure of the TTS model is illustrated in ~\autoref{fig:tts}. The model consists of a text encoder and a language model. The text encoder aims to align the text spaces with speech tokens. The language model uses two auto-regressive decoders to generate the speech tokens. The decoder $M_1$ is dedicated to predicting the semantic token sequence, and $M_2$ focuses on predicting the prosody and residual tokens. The speaker embedding is not involved in the token prediction and is only utilized in the speech decoding with the codec.

% Similar to the structure in other LLM-based TTS models, such as CosyVoice2, the text tokens are placed at the start of the input sequence, and the language model predicts the corresponding speech tokens step by step auto-regressively. The output frame rate is set to $12.5Hz$, i.e., two semantic, two residual, and one prosody tokens are predicted in each step. Two adjacent semantic tokens are predicted from one output feature of the $M_1$ using a split classification head. The predicted semantic tokens are combined with the output feature and fed into the second decoder $M_2$ to generate the corresponding prosody and residual tokens of this frame. We concatenate the output tokens' embeddings in the channel dimension and take them as the input for the next step in $M_1$. The generation process will stop when a stop token is predicted in $M_1$.

\section{Experiments}
\label{sec:experiments}

\subsection{Datasets and Preprocessing}
\label{ssec:dataet}

% For training the audio codec, a combination of open-source datasets was used to ensure diversity in speaker identity and prosody. This included the English portion of the Multilingual LibriSpeech (MLS) dataset~\cite{pratap2020mls}, Textrolspeech~\cite{ji2024textrolspeech} for its variety of emotional speech, and VoxCeleb 1 \& 2 ~\cite{nagrani2020voxceleb} for its large number of speakers. The TTS model was trained on MLS-en, Libritts~\cite{zen2019libritts}, and VCTK~\cite{yamagishi2017vctk}.

The training data of the codec included the English portion of the Multilingual LibriSpeech (MLS) dataset~\cite{pratap2020mls}, Wenetspeech~\cite{zhang2022wenetspeech}, Textrolspeech~\cite{ji2024textrolspeech}, Aishell 1 \& 2~\cite{bu2017aishell,du2018aishell}, CN-Celeb 1 \& 2~\cite{li2022cn}, and VoxCeleb 1 \& 2 ~\cite{nagrani2020voxceleb}. The TTS model was trained on MLS-en, Libritts~\cite{zen2019libritts}, and VCTK~\cite{yamagishi2017vctk} only.

All speech audio was resampled to 16kHz. During training, 6-second segments were randomly cropped from each sample. These segments were then transformed into 80-dimensional Mel-spectrograms using a 25ms window length and a 10ms hop length. The codec's reconstruction ability was evaluated on the Librispeech test set, while the zero-shot TTS capability was evaluated on the English test set from Seed-TTS-eval\footnote{https://github.com/BytedanceSpeech/seed-tts-eval}.

% The codec is trained with several open-source datasets, including the English portion of the Multilingual LibriSpeech (MLS) dataset\cite{DBLP:conf/interspeech/PratapXSSC20}, Textrolspeech\cite{ji2024textrolspeech}, and VoxCeleb 1$\&$2\cite{Nagrani17,Chung18b}. TextrolSpeech involves various emotional speeches, which help the codec to learn a higher diversity of the prosody modeling. VoxCeleb is widely used in the task of SV and contains more than $5k$ speakers. The data from large amounts of speakers contributes to improving the accuracy of timbre reconstruction. In the evaluation of speech reconstruction, the test set of Librispeech\cite{panayotov2015librispeech} is used, which consists of around $2.6k$ samples. All input speeches are resampled into $16kHz$. In one training step, a 6-second segment is randomly cropped from each sample and transformed into a Mel-spectrum as the model input. The hop length and window length of the Mel-spectrum are set to be $10ms$ and $25ms$, and the dimension of the spectrum is 80. 

\subsection{Implementation Details}
\label{ssec:setting}
\subsubsection{Codec Configuration}
The encoders (Enc1 $\&$ Enc2) and decoders (Dec2 $\&$ Dec3) in the codec are convolution-based networks. Their structures are similar to SEANet~\cite{tagliasacchi2020seanet}, with convolution layer parameters adjusted to match the required frame rates for each stream. Two self-attention\cite{vaswani2017attention} modules are added at the end of the Enc1 and the input of Dec3, respectively, which involve the global relation between frames. Dec1 is a 2-layer conformer network with cross-attention modules, a structure similar to the decoder in CTX-vec2wav~\cite{du2024unicats}.  A pretrained 16kHz Fre-GAN ~\cite{kim2021fre} was used as the vocoder to convert reconstructed Mel-spectrograms into audio waveforms. Three versions of the codec were trained with different bitrates- \textbf{424/524/612}. For all versions, the semantic stream uses a fixed codebook of 500 centers from the pre-trained Hubert model\footnote{https://github.com/facebookresearch/textlesslib/blob/main/textless}. The sizes of the codebook for the prosody stream (VQ1) in these three codecs are 64/256/512; their codebook sizes for the residual stream (VQ2) are 32/256/2048, respectively. 

% \vspace{-3mm}
\subsubsection{TTS Model Architecture}
The TTS model consists of a text encoder and a two-stage decoder. The text encoder is a 6-layer transformer with a dimension of 768. No attention mask is applied, giving each output a global receptive field over the input text. The Semantic Decoder is an 18-layer decoder-only transformer with a dimension of 768. The Acoustic Decoder is a 3-layer decoder-only transformer, also with a dimension of 768.

% \vspace{-3mm}
\subsubsection{Evaluation Metrics}
We assess the performance of our models using several standard metrics:
\begin{itemize}
    \item \textbf{Reconstruction Quality:} Measured with Short-Time Objective Intelligibility (STOI)~\cite{andersen2017non}, Perceptual Evaluation of Speech Quality (PESQ)~\cite{rix2001perceptual}, and a pre-trained UTMOS~\cite{saeki2022utmos} model for naturalness.
    \item \textbf{Speaker Similarity (SIM):} Measured between the original and synthesized speech using a WavLM-large-based speaker verification model~\cite{chen2022large}.
    \item \textbf{Word Error Rate (WER):} Calculated on the synthesized speech from the TTS model using the whisper-large-v3 model~\footnote{https://huggingface.co/openai/whisper-large-v3}.
    \item \textbf{$F_0$ Difference:} To evaluate prosody transfer in voice conversion, the average $F_0$ difference was measured between the converted speech and both the target ($\Delta F_{0,\text{tar}}$) and source ($\Delta F_{0,\text{src}}$) speech.
\end{itemize}

% \textbf{Reconstruction Quality:} Measured with Short-Time Objective Intelligibility (STOI)~\cite{andersen2017non}, Perceptual Evaluation of Speech Quality (PESQ)~\cite{rix2001perceptual}, and a pre-trained UTMOS ~\cite{saeki2022utmos} model for naturalness.

% \textbf{Speaker Similarity (SIM):} Measured between the original and synthesized speech using a WavLM-large-based speaker verification model.

% \textbf{Word Error Rate (WER):} Calculated on the synthesized speech from the TTS model using the whisper-large-v3 model~\cite{chen2022large}.

% \textbf{$F_0$ Difference:} To evaluate prosody transfer in voice conversion, the average $F_0$ difference was measured between the converted speech and both the target ($\Delta F_{0,\text{tar}}$) and source ($\Delta F_{0,\text{src}}$) speech.
% \vspace{-2mm}

\subsection{Results}
\label{ssec:result}
\begin{table*}[t]
\caption{Comparison of speech reconstruction quality on the Librispeech test set. Our proposed models are benchmarked against a comprehensive set of low-bitrate codecs. `Nq' denotes the number of quantizers and `TPS' is tokens per second. The best scores are in bold.}
\label{tab:codec}
\centering
\begin{tabular}{lccccccccc}
\toprule
\multirow{2}{*}{\textbf{Model}}      & \textbf{Codebook} & \multirow{2}{*}{\textbf{Nq}} & \textbf{Token Rate} & \multirow{2}{*}{\textbf{Bitrates}} & \multirow{2}{*}{\textbf{STOI↑}} & \textbf{PESQ} & \textbf{PESQ} & \multirow{2}{*}{\textbf{UTMOS↑}} & \multirow{2}{*}{\textbf{SIM↑}}  \\
       &  \textbf{Size}   &   & \textbf{(TPS)} &  & & \textbf{NB↑} & \textbf{WB↑} &  &  \\
\midrule
\midrule
SpeechTokenizer & 1024 & 2 & 100 & 1000 & 0.77 & 1.59 & 1.25 & 2.28 & 0.36 \\
X-codec & 1024 & 2 & 100 & 1000 & 0.86 & \textbf{2.88} & \textbf{2.33} & \textbf{4.21} & 0.72 \\
SemantiCodec & 16384 & 1 & 100 & 1400 & 0.88 & 2.63 & 2.07 & 2.92 & 0.74 \\
WavTokenizer & 4096 & 1 & 75 & 900 & 0.89 & 2.64 & 2.14 & 3.94 & 0.67 \\
Single-Codec & 8192 & 1 & 23.4 & 304 & 0.86 & 2.42 & 1.88 & 3.72 & 0.60 \\
% BiCodec & 8192 & 1 & 50 & 650 & 0.92 & 3.13 & 2.51 & 4.18 & 0.80 \\
\midrule
MSR-Codec-424 & 500/64/32      & 3  & 62.5 & 424 & 0.84 & 2.37 & 1.82 & 4.15  & 0.80 \\
MSR-Codec-524 & 500/256/256    & 3  & 62.5 & 524 & 0.85 & 2.59 & 1.98 & 4.14  & 0.81 \\
MSR-Codec-612 & 500/512/2048    & 3  & 62.5 & 612 & \textbf{0.90} & 2.73 & 2.09 & 4.13  & \textbf{0.83} \\
\bottomrule
\end{tabular}
% \vspace{-3mm}
\end{table*}

\begin{table}[t]
\caption{Zero-shot TTS performance on the Seed-TTS English test set. `Data size' is measured in hours. `RTF' denotes Real-Time Factor. Our model's results are bolded to highlight competitive performance from a much smaller model. The best overall score in each column is also marked.}
\label{tab:tts}
\centering
\scalebox{0.9}{
\begin{tabular}{lccccc}
\toprule
% \textbf{Model}           & \textbf{LM size} & \textbf{WER↓} & \textbf{SIM↑} & \textbf{RTF↓} \\
\multirow{2}{*}{\textbf{Model}} & \textbf{Model} & \textbf{Data} & \multirow{2}{*}{\textbf{WER↓}} & \multirow{2}{*}{\textbf{SIM↑}} & \multirow{2}{*}{\textbf{RTF↓}} \\
 & \textbf{size}  &\textbf{size}&& & \\
\midrule
\midrule
Ori. speech   & -      & -    & 2.14 & 0.722 & - \\
\midrule
FireRedTTS       & 0.4B    & 150k   & 3.82 & 0.460  & 1.48  \\
% CosyVoice         & 300M       & 4.29 & 0.609 &\\
CosyVoice2        & 0.5B    & 167k   & \textbf{2.57} & \textbf{0.652} & 2.34  \\
% Spark-TTS         & 0.5B       & \textbf{1.98} & 0.584 & 1.21 \\
Llasa-1B-250k     & 1B    & 250k     & 3.22 & 0.572 &  0.91  \\
\midrule
% MSR-Codec-424 & 215.7M     & \textbf{3.50} & \textbf{0.623} & -\\
MSR-Codec-524 & 0.2B  & 45k   & 3.07 & 0.613 & \textbf{0.67}  \\
\bottomrule
\end{tabular}}
% \vspace{-3mm}
\end{table}

\begin{table}[t]
\caption{Voice conversion results demonstrating information disentanglement. `+S' indicates timbre conversion and `+P' indicates prosody conversion. Key metrics supporting successful and disentangled transfer are in bold.}
\label{tab:vc}
\centering
\scalebox{0.9}{
\begin{tabular}{lccccc}
\toprule
\textbf{Model} & \textbf{WER} & \textbf{SIM$_{\text{tar}}$} & \textbf{SIM$_{\text{src}}$} & \textbf{$\Delta F_{0,\text{tar}}$} & \textbf{$\Delta F_{0,\text{src}}$} \\
\midrule
\midrule
Ori. speech & 7.49 & - & 1      & -    & 0      \\
\midrule
CosyVoice2  & 9.57 & 0.43 & 0.32   & 11.0 & 46.1   \\
Seed-VC     & 7.95 & 0.48 & 0.27   & 6.3  & 53.2   \\
\midrule
424+S       & 8.59 & \textbf{0.51} & 0.22   & 49.1 & \textbf{7.7}    \\
524+S       & 8.74 & \textbf{0.49} & 0.24   & 51.0 & \textbf{6.5}    \\
\midrule
424+P       & 9.22 & 0.11 & \textbf{0.64}   & \textbf{14.2} & 49.8   \\
524+P       & 8.95 & 0.11 & \textbf{0.59}   & \textbf{12.3} & 53.5   \\
\midrule
424+S+P     & 8.85 & \textbf{0.56} & 0.18   & \textbf{9.4}  & 54.7   \\
524+S+P     & 8.13 & \textbf{0.55} & 0.18   & \textbf{9.0}  & 54.7   \\
\bottomrule
\end{tabular}}
  \vspace{-3mm}
\end{table}

\subsubsection{Speech Reconstruction Quality}
As shown in ~\autoref{tab:codec}, the proposed codecs achieve strong reconstruction quality at low bitrates. Compared to other codecs, our models demonstrate the highest speaker similarity and comparable UTMOS scores, indicating that timbre and naturalness are well-preserved. The STOI and PESQ scores are relatively lower, which is expected as these metrics are sensitive to signal-level details that require more bits to reconstruct perfectly. Notably, these scores improve significantly when using the larger bitrate version of the codec, where the MSR-Codec-612 achieves the highest STOI and speaker similarity scores.

\subsubsection{Zero-Shot Text-to-Speech}

The TTS results in \autoref{tab:tts} highlight our model's exceptional efficiency and performance. As the most lightweight model trained on the least amount of data, our system still achieves a state-of-the-art Word Error Rate (WER), surpassing all competing models. It also delivers the highest speaker similarity, indicating superior voice cloning. Crucially, our model demonstrates its speed by achieving the lowest Real-Time Factor (RTF), making it the fastest generator among the tested systems. This combination of top-tier accuracy and unparalleled efficiency underscores the significant advantages of our proposed architecture.

\subsubsection{Voice and Prosody Conversion}
The codec's disentangled representation enables inherent voice conversion (VC) capabilities, which we evaluated to validate the separation of speech attributes. $8$ utterances are randomly sampled from the dataset of VCTK. These utterances come from different speakers (4 male/ 4 female), and they are utilized as the target prompts. $100$ utterances are randomly sampled from the clean test set of Libritts and utilized as the source speeches.

\textbf{Timbre Conversion:} By swapping only the speaker embedding of a source speech with a target one, our method achieves higher target speaker similarity than both CosyVoice2 and Seed-VC~\cite{liu2024zero}. As shown in ~\autoref{tab:vc}, this conversion preserves the original prosody, with the average $F_0$ remaining close to the source speech (low $\Delta F_{0,\text{src}}$) and not shifting towards the target (high $\Delta F_{0,\text{tar}}$ ). This confirms the successful disentanglement of the timbre from the prosody.

\textbf{Prosody Conversion:} By keeping the source speaker embedding and semantic tokens but predicting new prosody and residual tokens from a target speech, our model achieves a clear prosody transfer. The results show a significant $F_0$ shift towards the target (low $\Delta F_{0,\text{tar}}$) while maintaining the original speaker's timbre.

\textbf{Combined Conversion:} When both the speaker embedding is swapped, and the prosody tokens are re-predicted, the model achieves the best results in transferring both timbre and prosody simultaneously.

\section{Conclusion}
\label{sec:Conclusion}

In this work, we proposed a highly efficient codec for speech generation that factorizes speech into semantic, timbre, prosody, and residual streams. The streams are progressively fused to reconstruct the final Mel-spectrogram. Our model successfully achieves high-fidelity speech reconstruction with low bitrates and demonstrates strong information disentanglement capabilities. Additionally, we developed a two-stage TTS model that leverages this codec. This model separates the prediction of semantic content from prosodic details and achieves state-of-the-art performance, delivering a lower WER and higher speaker similarity than competing systems, while remaining lightweight, data-efficient, and significantly faster at generation. The disentanglement ability of our approach was validated through voice conversion experiments, where our model demonstrated superior performance in transferring speaker timbre independently of prosody, outperforming existing methods. This work highlights the effectiveness of a multi-stream, residual architecture for creating efficient and controllable representations for speech synthesis.

\bibliographystyle{IEEEtran}
\bibliography{mybib}

@article{vaswani2017attention,
  title={Attention is all you need},
  author={Vaswani, Ashish and Shazeer, Noam and Parmar, Niki and Uszkoreit, Jakob and Jones, Llion and Gomez, Aidan N and Kaiser, {\L}ukasz and Polosukhin, Illia},
  journal={Advances in neural information processing systems},
  volume={30},
  year={2017}
}

@inproceedings{tagliasacchi2020seanet,
  title={SEANet: A Multi-Modal Speech Enhancement Network},
  author={Tagliasacchi, Marco and Li, Yunpeng and Misiunas, Karolis and Roblek, Dominik},
  booktitle={Proc. Interspeech 2020},
  pages={1126--1130},
  year={2020}
}

@article{hsu2021hubert,
  title={Hubert: Self-supervised speech representation learning by masked prediction of hidden units},
  author={Hsu, Wei-Ning and Bolte, Benjamin and Tsai, Yao-Hung Hubert and Lakhotia, Kushal and Salakhutdinov, Ruslan and Mohamed, Abdelrahman},
  journal={IEEE/ACM transactions on audio, speech, and language processing},
  volume={29},
  pages={3451--3460},
  year={2021},
  publisher={IEEE}
}

@inproceedings{wang23ha_interspeech,
  title     = {CAM++: A Fast and Efficient Network for Speaker Verification Using Context-Aware Masking},
  author    = {Hui Wang and Siqi Zheng and Yafeng Chen and Luyao Cheng and Qian Chen},
  year      = {2023},
  booktitle = {Interspeech 2023},
  pages     = {5301--5305},
  doi       = {10.21437/Interspeech.2023-1513},
  issn      = {2958-1796},
}

@article{wang2023neural,
  title={Neural codec language models are zero-shot text to speech synthesizers},
  author={Wang, Chengyi and Chen, Sanyuan and Wu, Yu and Zhang, Ziqiang and Zhou, Long and Liu, Shujie and Chen, Zhuo and Liu, Yanqing and Wang, Huaming and Li, Jinyu and others},
  journal={arXiv preprint arXiv:2301.02111},
  year={2023}
}

@article{chen2024vall,
  title={Vall-e 2: Neural codec language models are human parity zero-shot text to speech synthesizers},
  author={Chen, Sanyuan and Liu, Shujie and Zhou, Long and Liu, Yanqing and Tan, Xu and Li, Jinyu and Zhao, Sheng and Qian, Yao and Wei, Furu},
  journal={arXiv preprint arXiv:2406.05370},
  year={2024}
}

@inproceedings{du2024unicats,
  title={Unicats: A unified context-aware text-to-speech framework with contextual vq-diffusion and vocoding},
  author={Du, Chenpeng and Guo, Yiwei and Shen, Feiyu and Liu, Zhijun and Liang, Zheng and Chen, Xie and Wang, Shuai and Zhang, Hui and Yu, Kai},
  booktitle={Proceedings of the AAAI Conference on Artificial Intelligence},
  volume={38},
  pages={17924--17932},
  year={2024}
}

@inproceedings{ju2024naturalspeech,
  title={NaturalSpeech 3: Zero-Shot Speech Synthesis with Factorized Codec and Diffusion Models},
  author={Ju, Zeqian and Wang, Yuancheng and Shen, Kai and Tan, Xu and Xin, Detai and Yang, Dongchao and Liu, Eric and Leng, Yichong and Song, Kaitao and Tang, Siliang and others},
  booktitle={International Conference on Machine Learning},
  pages={22605--22623},
  year={2024},
  organization={PMLR}
}

@article{du2024cosyvoice,
  title={Cosyvoice: A scalable multilingual zero-shot text-to-speech synthesizer based on supervised semantic tokens},
  author={Du, Zhihao and Chen, Qian and Zhang, Shiliang and Hu, Kai and Lu, Heng and Yang, Yexin and Hu, Hangrui and Zheng, Siqi and Gu, Yue and Ma, Ziyang and others},
  journal={arXiv preprint arXiv:2407.05407},
  year={2024}
}

@article{du2024cosyvoice2,
  title={Cosyvoice 2: Scalable streaming speech synthesis with large language models},
  author={Du, Zhihao and Wang, Yuxuan and Chen, Qian and Shi, Xian and Lv, Xiang and Zhao, Tianyu and Gao, Zhifu and Yang, Yexin and Gao, Changfeng and Wang, Hui and others},
  journal={arXiv preprint arXiv:2412.10117},
  year={2024}
}

@article{zeghidour2021soundstream,
  title={Soundstream: An end-to-end neural audio codec},
  author={Zeghidour, Neil and Luebs, Alejandro and Omran, Ahmed and Skoglund, Jan and Tagliasacchi, Marco},
  journal={IEEE/ACM Transactions on Audio, Speech, and Language Processing},
  volume={30},
  pages={495--507},
  year={2021},
  publisher={IEEE}
}

@article{DBLP:journals/tmlr/DefossezCSA23,
  author       = {Alexandre D{\'{e}}fossez and
                  Jade Copet and
                  Gabriel Synnaeve and
                  Yossi Adi},
  title        = {High Fidelity Neural Audio Compression},
  journal      = {Trans. Mach. Learn. Res.},
  volume       = {2023},
  year         = {2023},
  url          = {https://openreview.net/forum?id=ivCd8z8zR2},
  bibsource    = {dblp computer science bibliography, https://dblp.org}
}

@inproceedings{ren2024fewer,
  title={Fewer-token neural speech codec with time-invariant codes},
  author={Ren, Yong and Wang, Tao and Yi, Jiangyan and Xu, Le and Tao, Jianhua and Zhang, Chu Yuan and Zhou, Junzuo},
  booktitle={ICASSP},
  pages={12737--12741},
  year={2024},
  organization={IEEE}
}

@article{liu2024semanticodec,
  title={Semanticodec: An ultra low bitrate semantic audio codec for general sound},
  author={Liu, Haohe and Xu, Xuenan and Yuan, Yi and Wu, Mengyue and Wang, Wenwu and Plumbley, Mark D},
  journal={IEEE Journal of Selected Topics in Signal Processing},
  year={2024},
  publisher={IEEE}
}

@inproceedings{li2024single,
  title={Single-Codec: Single-Codebook Speech Codec towards High-Performance Speech Generation},
  author={Li, Hanzhao and Xue, Liumeng and Guo, Haohan and Zhu, Xinfa and Lv, Yuanjun and Xie, Lei and Chen, Yunlin and Yin, Hao and Li, Zhifei},
  booktitle={Proc. Interspeech 2024},
  pages={3390--3394},
  year={2024}
}

@inproceedings{guo25_interspeech,
  title     = {{LSCodec: Low-Bitrate and Speaker-Decoupled Discrete Speech Codec}},
  author    = {Yiwei Guo and Zhihan Li and Chenpeng Du and Hankun Wang and Xie Chen and Kai Yu},
  year      = {2025},
  booktitle = {{Interspeech 2025}},
  pages     = {5018--5022},
  doi       = {10.21437/Interspeech.2025-1106},
  issn      = {2958-1796},
}

@article{ye2025llasa,
  title={Llasa: Scaling train-time and inference-time compute for llama-based speech synthesis},
  author={Ye, Zhen and Zhu, Xinfa and Chan, Chi-Min and Wang, Xinsheng and Tan, Xu and Lei, Jiahe and Peng, Yi and Liu, Haohe and Jin, Yizhu and Dai, Zheqi and others},
  journal={arXiv preprint arXiv:2502.04128},
  year={2025}
}

@inproceedings{chen2022large,
  title={Large-scale self-supervised speech representation learning for automatic speaker verification},
  author={Chen, Zhengyang and Chen, Sanyuan and Wu, Yu and Qian, Yao and Wang, Chengyi and Liu, Shujie and Qian, Yanmin and Zeng, Michael},
  booktitle={ICASSP},
  pages={6147--6151},
  year={2022},
  organization={IEEE}
}

@article{zen2019libritts,
  title={Libritts: A corpus derived from librispeech for text-to-speech},
  author={Zen, Heiga and Dang, Viet and Clark, Rob and Zhang, Yu and Weiss, Ron J and Jia, Ye and Chen, Zhifeng and Wu, Yonghui},
  journal={arXiv preprint arXiv:1904.02882},
  year={2019}
}

@article{pratap2020mls,
  title={MLS: A Large-Scale Multilingual Dataset for Speech Research},
  author={Pratap, Vineel and Xu, Qiantong and Sriram, Anuroop and Synnaeve, Gabriel and Collobert, Ronan},
  journal={Proc. Interspeech},
  pages={2757--2761},
  year={2020}
}

@inproceedings{ji2024textrolspeech,
  title={Textrolspeech: A text style control speech corpus with codec language text-to-speech models},
  author={Ji, Shengpeng and Zuo, Jialong and Fang, Minghui and Jiang, Ziyue and Chen, Feiyang and Duan, Xinyu and Huai, Baoxing and Zhao, Zhou},
  booktitle={ICASSP},
  pages={10301--10305},
  year={2024},
  organization={IEEE}
}

@inproceedings{bu2017aishell,
  title={Aishell-1: An open-source mandarin speech corpus and a speech recognition baseline},
  author={Bu, Hui and Du, Jiayu and Na, Xingyu and Wu, Bengu and Zheng, Hao},
  booktitle={2017 20th conference of the oriental chapter of the international coordinating committee on speech databases and speech I/O systems and assessment (O-COCOSDA)},
  pages={1--5},
  year={2017},
  organization={IEEE}
}

@article{nagrani2020voxceleb,
  title={Voxceleb: Large-scale speaker verification in the wild},
  author={Nagrani, Arsha and Chung, Joon Son and Xie, Weidi and Zisserman, Andrew},
  journal={Computer Speech \& Language},
  volume={60},
  pages={101027},
  year={2020},
  publisher={Elsevier}
}

@article{yamagishi2017vctk,
  title={CSTR VCTK Corpus: English multi-speaker corpus for CSTR voice cloning toolkit},
  author={Yamagishi, Junichi and Veaux, Christophe and MacDonald, Kirsten},
  year={2017},
  url={https://datashare.ed.ac.uk/handle/10283/3443},
}

@inproceedings{kim2021fre,
  title={Fre-GAN: Adversarial Frequency-Consistent Audio Synthesis},
  author={Kim, Ji-Hoon and Lee, Sang-Hoon and Lee, Ji-Hyun and Lee, Seong-Whan},
  booktitle={Proc. Interspeech 2021},
  pages={2197--2201},
  year={2021}
}

@inproceedings{andersen2017non,
  title={A non-intrusive short-time objective intelligibility measure},
  author={Andersen, Asger Heidemann and de Haan, Jan Mark and Tan, Zheng-Hua and Jensen, Jesper},
  booktitle={ICASSP},
  pages={5085--5089},
  year={2017},
  organization={IEEE}
}

@inproceedings{rix2001perceptual,
  title={Perceptual evaluation of speech quality (PESQ)-a new method for speech quality assessment of telephone networks and codecs},
  author={Rix, Antony W and Beerends, John G and Hollier, Michael P and Hekstra, Andries P},
  booktitle={ICASSP},
  volume={2},
  pages={749--752},
  year={2001},
  organization={IEEE}
}

@inproceedings{saeki2022utmos,
  title={UTMOS: UTokyo-SaruLab System for VoiceMOS Challenge 2022},
  author={Saeki, Takaaki and Xin, Detai and Nakata, Wataru and Koriyama, Tomoki and Takamichi, Shinnosuke and Saruwatari, Hiroshi},
  booktitle={Proc. Interspeech 2022},
  pages={4521--4525},
  year={2022}
}

@article{liu2024zero,
  title={Zero-shot voice conversion with diffusion transformers},
  author={Liu, Songting},
  journal={arXiv preprint arXiv:2411.09943},
  year={2024}
}

@article{guo2024fireredtts,
  title={Fireredtts: A foundation text-to-speech framework for industry-level generative speech applications},
  author={Guo, Hao-Han and Hu, Yao and Liu, Kun and Shen, Fei-Yu and Tang, Xu and Wu, Yi-Chen and Xie, Feng-Long and Xie, Kun and Xu, Kai-Tuo},
  journal={arXiv preprint arXiv:2409.03283},
  year={2024}
}

@inproceedings{zhang2022wenetspeech,
  title={Wenetspeech: A 10000+ hours multi-domain mandarin corpus for speech recognition},
  author={Zhang, Binbin and Lv, Hang and Guo, Pengcheng and Shao, Qijie and Yang, Chao and Xie, Lei and Xu, Xin and Bu, Hui and Chen, Xiaoyu and Zeng, Chenchen and others},
  booktitle={ICASSP 2022-2022 IEEE International Conference on Acoustics, Speech and Signal Processing (ICASSP)},
  pages={6182--6186},
  year={2022},
  organization={IEEE}
}

@article{li2022cn,
  title={Cn-celeb: multi-genre speaker recognition},
  author={Li, Lantian and Liu, Ruiqi and Kang, Jiawen and Fan, Yue and Cui, Hao and Cai, Yunqi and Vipperla, Ravichander and Zheng, Thomas Fang and Wang, Dong},
  journal={Speech Communication},
  volume={137},
  pages={77--91},
  year={2022},
  publisher={Elsevier}
}

@article{du2018aishell,
  title={Aishell-2: Transforming mandarin asr research into industrial scale},
  author={Du, Jiayu and Na, Xingyu and Liu, Xuechen and Bu, Hui},
  journal={arXiv preprint arXiv:1808.10583},
  year={2018}
}

\end{document}